# Impact of Artificial Intelligence on Businesses: from Research, Innovation, Market Deployment to Future Shifts in Business Models[1]

Neha Soni[1], Enakshi Khular Sharma[1], Narotam Singh[2], Amita Kapoor[3],

[1]Department of Electronic Science, University of Delhi South Campus, Delhi, India
[2]Information Communication and Instrumentation Training Centre, India Meteorological Department, Ministry of Earth Sciences, Delhi, India
[3]Shaheed Rajguru College of Applied Sciences for Women, University of Delhi, Delhi, India
{soni.neha2191@gmail.com, enakshi54@yahoo.co.in, narotam.singh@gmail.com, dr.amita.kapoor@ieee.org}

## Abstract

The fast pace of artificial intelligence (AI) and automation is propelling strategists to reshape their business models. This is fostering the integration of AI in the business processes but the consequences of this adoption are underexplored and needs attention. This paper focuses on the overall impact of AI on businesses - from research, innovation, market deployment to future shifts in business models. To access this overall impact, we design a three dimensional research model, based upon the Neo-Schumpeterian economics and its three forces viz. innovation, knowledge, and entrepreneurship. The first dimension deals with research and innovation in AI. In the second dimension, we explore the influence of AI on the global market and the strategic objectives of the businesses and finally the third dimension examines how AI is shaping business contexts. Additionally, the paper explores AI implications on actors and its dark sides.

## Keywords

Artificial Intelligence, Automation, Digitization, Business Strategies, Innovation, Business Contexts

## 1. Introduction

The emerging technologies viz. internet of things (IoT), data science, big data, cloud computing, artificial intelligence (AI), and blockchain are changing the way we live, work and amuse ourselves. Further advancement of these technologies can contribute in developing hyper automation and hyper connectivity, which would bring us at the dawn of the Fourth Industrial Revolution or Industry 4.0 (Schwab 2017; Bloem 2014; Klosters 2016; Park 2017). Primarily, the advancement in AI is the heart of the enhanced performance of all other technologies and the evolution of Industry 4.0. This technological advancement, attributed to AI, would facilitate human-to-machine interactions, change the logic of business models, and transform the lifestyle and living standards of the human.

[1] Final manuscript is submitted to Journal of Business Research - Elsevier for consideration.

The adoption of AI is resulting in a world which is smarter and innovative. Route and traffic mapping by Google maps, price estimation of rides by Uber and Lyft, friends' tag suggestions at Facebook, spam filters in our email, recommendation for online shopping and cancer detection are only a few examples of AI technological innovations simplifying our lives. The incredible speed with which AI is entering every sector is forcing companies to get into the race to make their company an AI company. This is also impelling business, strategists, pioneers, entrepreneurs and investigators to use AI to design new strategies and create new sources of business value. Andrew Ng, co-founder Google Brain; former vice president & chief scientist of Baidu; co-chairman and co-founder of Coursera and an adjunct professor at Stanford University said in 2017 at Stanford MSx (Master of Science in Management for Experienced Leaders) Program (Lynch 2017)-

"Just as electricity transformed almost everything 100 years ago, today I actually have a hard time thinking of an industry that I don't think AI will transform in the next several years".

The statement carries considerable weight as Prof. Ng is one of the six top thinkers in AI and machine learning (Marr 2017), a prominent computer scientist and an AI entrepreneur i.e. he has the expertise in academia as well as industry. Therefore, it is necessary to pay attention to the wide-ranging implications of AI on governments, communities, companies, and individuals. This paper will address the influential academic achievements and innovations in the field of AI, its influence on the global market and the strategic objectives of organizations, and the shaping of business contexts with AI.

Research plays an increasingly important role in the development of innovations and new technology. Research and innovation process results in economic growth by encouraging the development of new markets and improving existing markets. According to the new growth economics, Neo-Schumpeterian Economics, there are three major forces which drive the economic dynamics: innovation, knowledge, and entrepreneurship (Hanusch 2006). Novelty and uncertainty is the most distinguishing mark of Neo-Schumpeterian Economics; the most visible form of novelty is innovation. Therefore, innovation (in particular, technological innovation) is the first major driving force of economic dynamics. The root of Neo-Schumpeterian Economics is to learn and search experimentally in permanently changing environments; thus knowledge (in particular, scientific knowledge) is considered to be the second major driving force of economic dynamics. Finally, the emphasis has been given on an entrepreneur; an economic actor who kicks off economic development by introducing novelties. Thus, according to Neo-Schumpeterian Economics, innovation, knowledge, and entrepreneurship are the three major driving forces of economic dynamics wherein the emergence of the new industries is driven by innovation, supported by knowledge and tested by the entrepreneurial action. (Hanusch 2006).

Based upon the above three forces (innovation, knowledge, and entrepreneurship), we design a three dimensional research model to access the overall impact of AI on businesses. The first dimension of our three dimensional research model, research and innovation in AI, explore the success of AI algorithms and investigate their deployment via commercially available intelligent machines and services. The second dimension, influence of automation and AI investigates the top AI organizations (companies and start-ups) and their strategic initiatives to deploy/launch AI-based services in existing and new industries. The third dimension, shaping of business contexts with AI, explores how the AI is transforming/ disrupting the regular flow of business activities.

The first dimension of our analysis was chosen to be the theory and algorithms behind AI-driven systems. For this, we needed information regarding recent developments in the field of data generation and processing, learning algorithms, and successful AI applications. To obtain this information we studied a large number of research papers published in reputed journals and conferences in the last fifteen years (2003-2018), explored various dataset and their providers, and identified commercially available intelligent machines and services. The results obtained from this analysis are summarized in a table (Table 1). The success and hype generated by the AI algorithms leads to the technological transformation in the global market. This transformation is forcing many existing companies to shift to AI and at the same time spawning a plethora of AI-based start-ups. Thus, in the second dimension, we identified the AI-motivated strategic initiatives of the corporate firms to grow faster with the most advanced technology of AI. We performed financial analysis of some of the top companies viz. Google, Apple, Amazon, Microsoft, and IBM working in AI and related fields. Apart from IBM, all these companies are founded within the last four decades and have managed to achieve top ranking at the global platform. To get a better understanding of the impact of AI on businesses we extended our analysis to include top AI start-ups in 2017 and 2018. We scavenged a large number of research blogs, recent reputed conferences, their sponsors, AI start-up acquisitions, market intelligence reports, stock market websites, company websites and official blogs to select the above AI companies (and start-ups) and perform their analysis. The integration of AI technology in the business processes results in the reshaping of business context, thus, in the third dimension we explored the influence of AI on the business contexts. The relevant data were extracted from business press releases, annual reports of the companies, innovation trend reports and research reports released by the companies providing market intelligence such as Gartner, Forrester, and IDC etc. From our analysis, we identified three business contexts that are majorly affected due to the adoption of AI-driven systems viz. Customer Interaction, Sales Platform, and Employee Skill Set.

## 2. Research Objectives

The inferences obtained from the above three-dimensional analysis will provide a better understanding of the innovations, the actual current degree of integration, application and the impact of AI in businesses. In conclusion, the above analysis provides the answers to the following questions:

- AI is a 60-year-old technology yet couldn't influence the society till last decade, what are the factors which are resulting in today's AI exponential growth?
- How "intelligent" machines and services are related to AI? Which of them are available for commercial use?
- What is behind all these real-world intelligent applications? Which AI algorithms are making these artificial systems intelligent?
- How is the growth of AI influencing all industries and sectors across the globe? Which countries are leading this race of AI?
- How the expansion of the technology in AI-enabled countries can lead to AI-divide – The 'dark side' to AI?
- Is this growth disrupting conventional business process? How is this influence of AI in myriad sectors transforming the market and the future jobs?

The answers to the above questions will help the human society to get prepared for the future challenges and accept the rapid changes occurring with the infusion of AI in human life and business. The present work is organized as follows: Sect. 2 focuses on the literature review of the sixty years of AI and today's reality, Sect. 3 provides the results obtained from the first-dimensional analysis i.e. the state-of-the-art (SOTA) research and innovations in AI;

Sect. 4 illustrates the results obtained from the second-dimensional analysis i.e. identification of strategic objectives, global market analysis of top AI companies and start-ups, Sect. 5 provides the third-dimensional analysis i.e. shaping of business contexts and finally some conclusions and directions for future research.

## 3. State-of-the-art of AI

The 63-year journey of AI has not been smooth, the period of hypes was followed by periods with reduced funding (also known as AI winters). However, despite these setbacks today we have AI back in limelight due to development of 'deep learning' neural networks with many hidden layers. This advancement of AI is attributed to two major factors: the availability of a large amount of data (big data), and hardware accelerators (graphics processing units (GPUs) and tensor processing units (TPUs)) (Goodfellow 2018; Abadi 2016). We investigated the popularity of AI on the web, news, and business in the last decade (June 2008 to December 2017) via Google Trends and CB Insights and the results have been illustrated in Fig.1. The Fig. 1 shows that from the year 2016, there has been a sudden increase in the number of mentions of the word "artificial intelligence" in business news and its search frequency on the web.

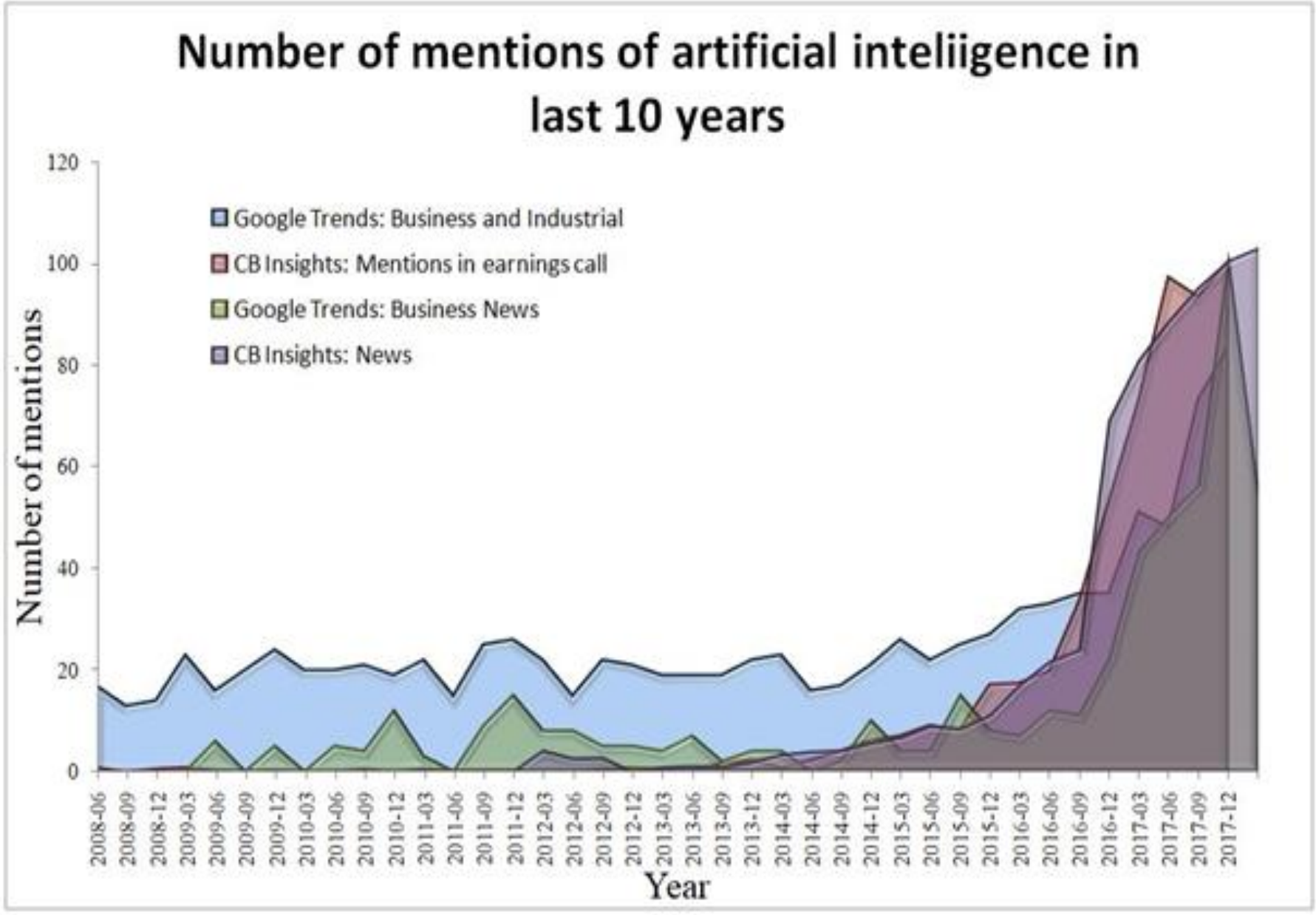

**Fig. 1** Number of mentions of “artificial intelligence” in the last 10 years

The increase in popularity of AI has led to an expansion in the investment in multiple sectors of AI including research, development, marketing, and production and vice versa. Some of the corporations with the maximum number of mentions of the word “artificial intelligence” in their earning calls are NVIDIA, ORBCOMM, Microsoft, and Facebook. These companies are making the technology commercially available in the form of APIs (application program interface), deep learning libraries, personal and professional agents, chatbots, robots, and many other exciting products. This is enhancing their business valuation, adding new dimensions of resources and making their products and services intelligent. At the same time, this AI-motivated paradigm shift is enhancing the mechanical, analytical, and intuitive skills of the employees and shaping other contexts of the businesses. According to International Data Corporation (IDC), the worldwide spending on cognitive and AI systems will increase prominently from $12 billion in 2017 to approx. $58 billion in 2021 (Columbus 2018; Shirer 2018). With this increased funding amount, AI is likely to involve in more sectors. Thus, it becomes necessary to investigate the functioning of an AI-driven system in different possible sectors in the existing businesses. At the same time, it is also important to inspect the market leaders and the start-ups adopting these AI-driven systems. This AI adoption by the companies will lead to shaping of the business contexts. Prior knowledge about all these domains will make the society aware about the development and adoption of AI in near future. This will also aid the policymakers in identifying challenges and exploring the legal and ethical corners in the field of AI. In the following section, we discuss the methodology adopted to investigate deeper about these domains and thus evaluate the overall impact of AI on businesses.

## 4. AI: Reshaping the Innovation Process

AI has reached a place where it can take real-world financial decisions, chat with people, play games against humans, and work hand in hand with them. Behind all these real-world applications, there is an AI-driven system or an intelligent agent (IA). It interacts with the environment in a repetitive cycle of sense-think-and-act. It takes in the data from the environment, makes an informed decision based on the input data and past experience, and finally performs an action affecting the environment. This IA can be a machine (industrial and home robots, self-driving cars) or a software agent (chatbots, recommender systems). It takes the data in the form of images, videos, sound, text etc., analyses this data using AI algorithms and delivers AI-powered solutions. Fig. 2 shows the sense-think-and-act cycle for an IA.

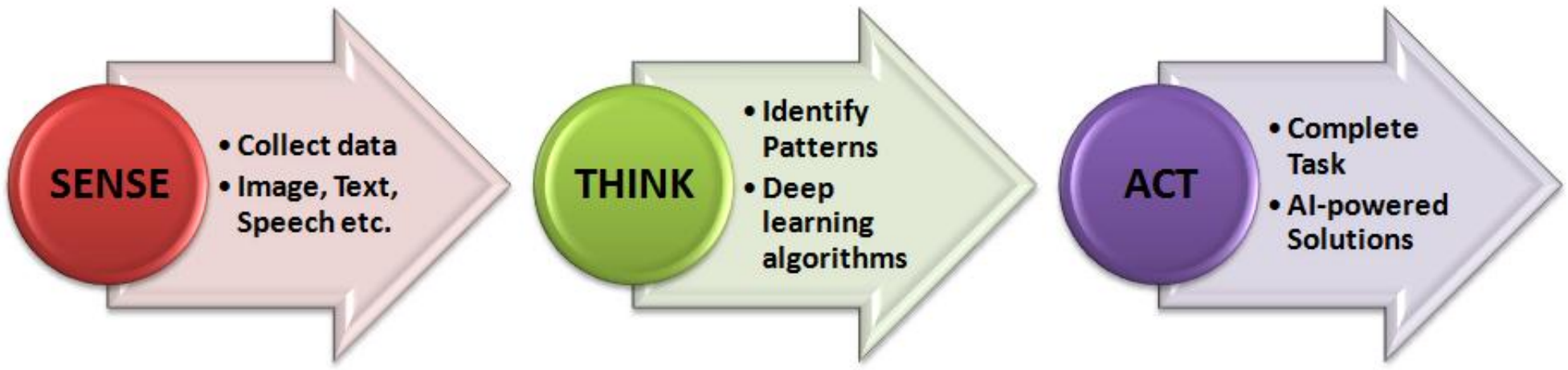

**Fig. 2** Sense-Think-and-Act process followed by an intelligent agent

### 4.1 Data: Fuel for the AI-Driven Systems

The unprecedented volume of data is the fuel for an AI-driven system. According to Satya Nadella, CEO, Microsoft "The core currency of any business will be the ability to convert their data into AI that drives competitive advantage". Earlier, the unavailability of data was hindering the progress of AI, however, in the last few years, the accessibility of low-cost and low-power sensors has resulted in the generation of a large amount of data. The data from the sensors like camera, global positioning system (GPS) unit, health monitoring sensor, smoke sensor, the chemical sensor, motion detection sensor etc. can be continuously streamed and processed on-the-fly or can be stored to gain useful insights by various mechanisms.

Moreover, the data from multiple sensors can be combined together using the technique of sensor fusion. There also exist other sources viz. online directories, review sites, surveys, actual retail sales, census database of countries, manually created/ hand-crafted data, e-commerce websites, online communities, etc. from where the data can be extracted using web scraping or other techniques. The raw data obtained from all these sources can be processed and used to train an IA; the conversion from raw data to processed data is an expensive and time-consuming task. A large number of online dataset sources exist, from where one can download different types of data.

Once an intelligent agent is trained, it can teach other agents and make them smarter. According to Professor Hod Lipson, Columbia University, this can be called as – "Machine teaching, one of the biggest exponential trends in AI" (Frank 2018). According to him "Data is the fuel of machine learning, but even for machines, some data is hard to get—it may be risky, slow, rare, or expensive. In those cases, machines can share experiences or create synthetic experiences for each other to augment or replace data. It turns out that this is not a minor effect, it actually is self-amplifying, and therefore exponential." (Frank 2018). The "machine teaching" will save human time and power as for every small change this would eliminate the need to train an agent from the scratch. This knowledge transfer within the agents would increase the development and deployment of IAs at a much faster rate.

## 4.2 Intelligent Thinking and Action Delivery: Algorithms and Output of the AI-Driven Systems

An IA explores the input data in order to learn correlations, detect similarities, extract features, and discover good representation at multiple levels. This requires the use of AI or machine learning tools like Bayesian algorithms, support vector machines (SVM), decision trees, deep learning networks (DLN) and ensemble configurations. Among these, the DLNs have emerged as the most popular approach in the last few years. Some of the DLNs have reported surpassing human-level accuracy in certain tasks. We explored a large number of research papers and books (Gulli 2017; Singh 2015; Krizhevsky 2012; Karpathy 2014; Simard 2003; Taigman 2014; He 2015; Vinyals 2015; Soni 2016; 2018b; Bahdanau 2014; Sutskever 2014, Wen 2015; Xiong 2017; Sak 2015; Amodei 2015; Chung 2011) published in reputed AI and machine learning journals and conferences such as Nature – International Journal of Science, Elsevier – Artificial Intelligence, Neural Information Processing System (NIPS), Computer Vision and Pattern Recognition (CVPR) etc. From the exploration of research papers, we found that for visual data convolutional neural networks (CNN) are preferred (Krizhevsky 2012; Karpathy 2014; Simard 2003; Taigman 2014; He 2015; Vinyals 2015). For textual data, the two variants of gated recurrent neural networks (gated-RNNs) i.e. LSTM (long short-term memory) and GRU (gated recurrent units) have been reported to give the best performance (Bahdanau 2014; Sutskever 2014, Wen 2015). For speech data, CNNs, RNNs and their combination have been preferred (Xiong 2017; Sak 2015; Amodei 2015; Chung 2011). More than human-level accuracy has been reported for certain tasks like human face recognition (Taigman 2014), traffic signal recognition (Cireşan 2015), MNIST data classification (Cireşan 2015), ImageNet data classification (He 2015), English language and Chinese (Mandarin) language transcription (Amodei 2015), breast cancer detection (Liu 2017) and Chinese language to English language news translation (Hassan 2018). Not only that, but the intelligent agents have also outperformed humans in the games like Go (Silver 2016), Chess (David 2016) and Atari (Mnih 2015).

The success and the hype generated by DLNs in the last few years have propelled many companies to launch a large number of AI-based machines and services. We identified some of the companies investing in AI; gathered the appropriate information about their customers, acquisitions, products, and services; analyzed this information to gain knowledge about the commercial availability of intelligent machines and services. Table 1 summarises the commercially available machines and services, the intelligent software behind them, their developers and number of people using these services worldwide.

**Table 1** commercially available intelligent machine and services, the intelligent software behind them, their developers and their user base.

This section provides ample evidence that AI has attained significant growth in terms of research, innovation, and deployment. One of the most notable achievements is surpassing the human-level accuracy in various tasks viz. games, recognition and classification. This offers various opportunities for process innovation and product innovation but issues like bias, trust, privacy and security still needs attention [Bostrom 2014, Etzioni 2017]. These are some of the issues which are compelling the researchers to think of the negative impact of AI. The next section explores the key trends in AI and the actual degree of its integration in various industries in the global market.

## 5. Influence of Research and Innovation in Automation and AI

### 5.1 On the Strategic Objectives of Organizations

The achievements in the field of research and innovation have propelled many existing companies to become AI companies and have spawned a plethora of AI-based start-ups. This has also increased the involvement of actors in AI-related research and academic events. For instance, one of the largest AI conferences, NIPS witnessed a 750% increase in the number of sponsors in a span of nine years (2010 to 2018). This exponential increase indicates the avidity of corporate firms to streamline with the recent research. Furthermore, actors are also adopting various other business strategies for their growth in the field of AI viz. recruiting AI talent, investing in core-AI companies and acquiring AI start-ups. According to an analysis by CB Insights, the number of AI start-up acquisitions increased to 422% in the last five years (2013 to 2017). This phenomenal increase indicates that corporate firms aim to grow faster with the most advanced technology of AI.

### 5.2 On the Global Market: Top Companies and Start-ups

In this section, first, we identify the top AI companies and then perform their financial analysis to assess the impact of automation and AI on the global market. We identified the top five AI companies from a list of 119+ corporate groups which participated in NIPS 2018. The criterion adopted for evaluation is the number of acquisitions of AI start-ups in the last 9 years (January 2010-January 2019). Fig. 3 shows the number of AI start-up acquisitions by the top five AI companies viz. Google, Apple, Amazon, Microsoft, and IBM. Table 2 summarises the details of the acquisitions and NIPS sponsorship level of the top five AI companies. In the following section, we perform the financial analysis of the top five AI companies viz. Google, Apple, Amazon, Microsoft, and IBM.

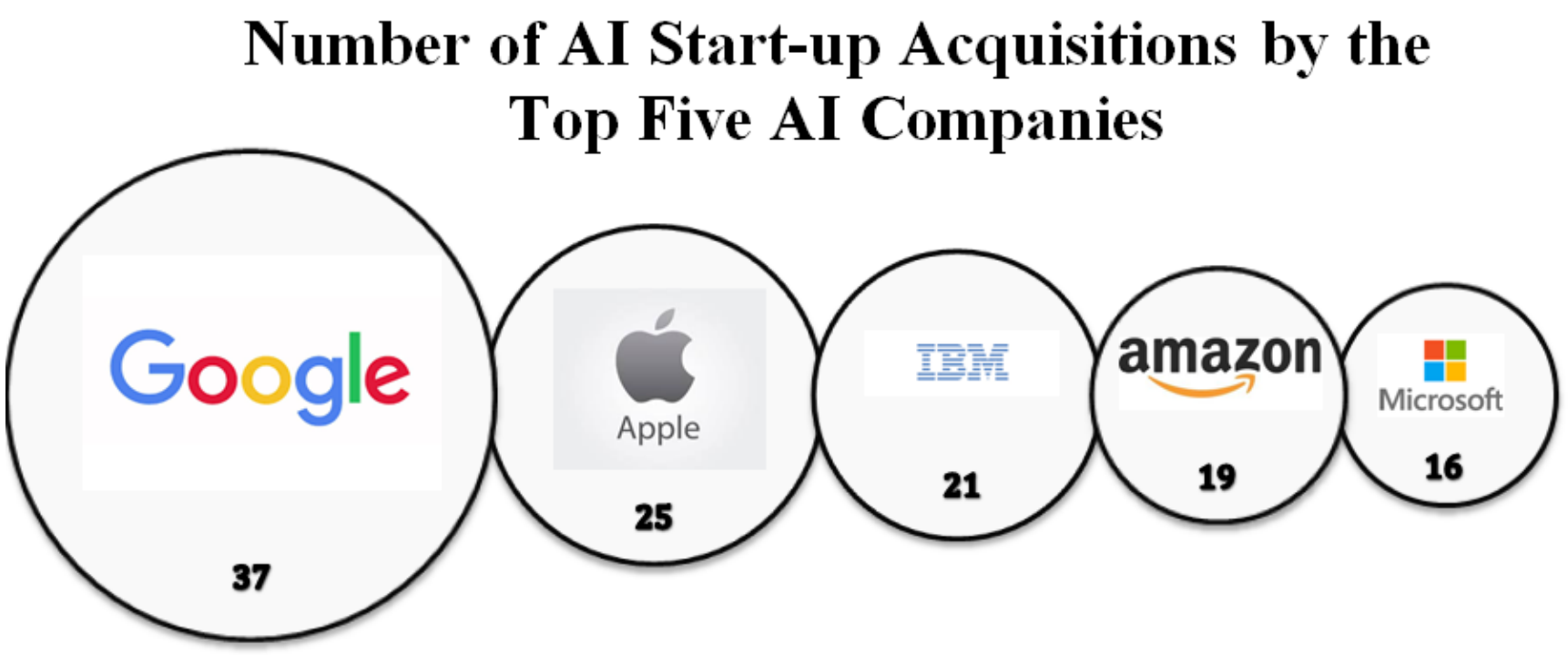

**Fig. 3** The number of AI start-up acquisitions by the top five AI companies

**Table 2** The details of the acquisitions and NIPS sponsorship level of the top five AI companies.

We performed the financial analysis of the top AI companies by analyzing their financial worth in the last decade. Our analysis shows an increasing trend in the share prices, EPS (earning per share), investment in AI, and the net sales of all the top companies for the last decade (2009-2018). Fig 4 shows the trend of normalized net sales of the top five AI companies for the last 10 years (2009-2018). Apart from Apple and IBM, all other companies show a continual increase over the years. The investment done in AI must have played some role in the financial growth of these companies, but it was not possible to conclusively find a direct correspondence between the two because there can be various factors behind the growth.

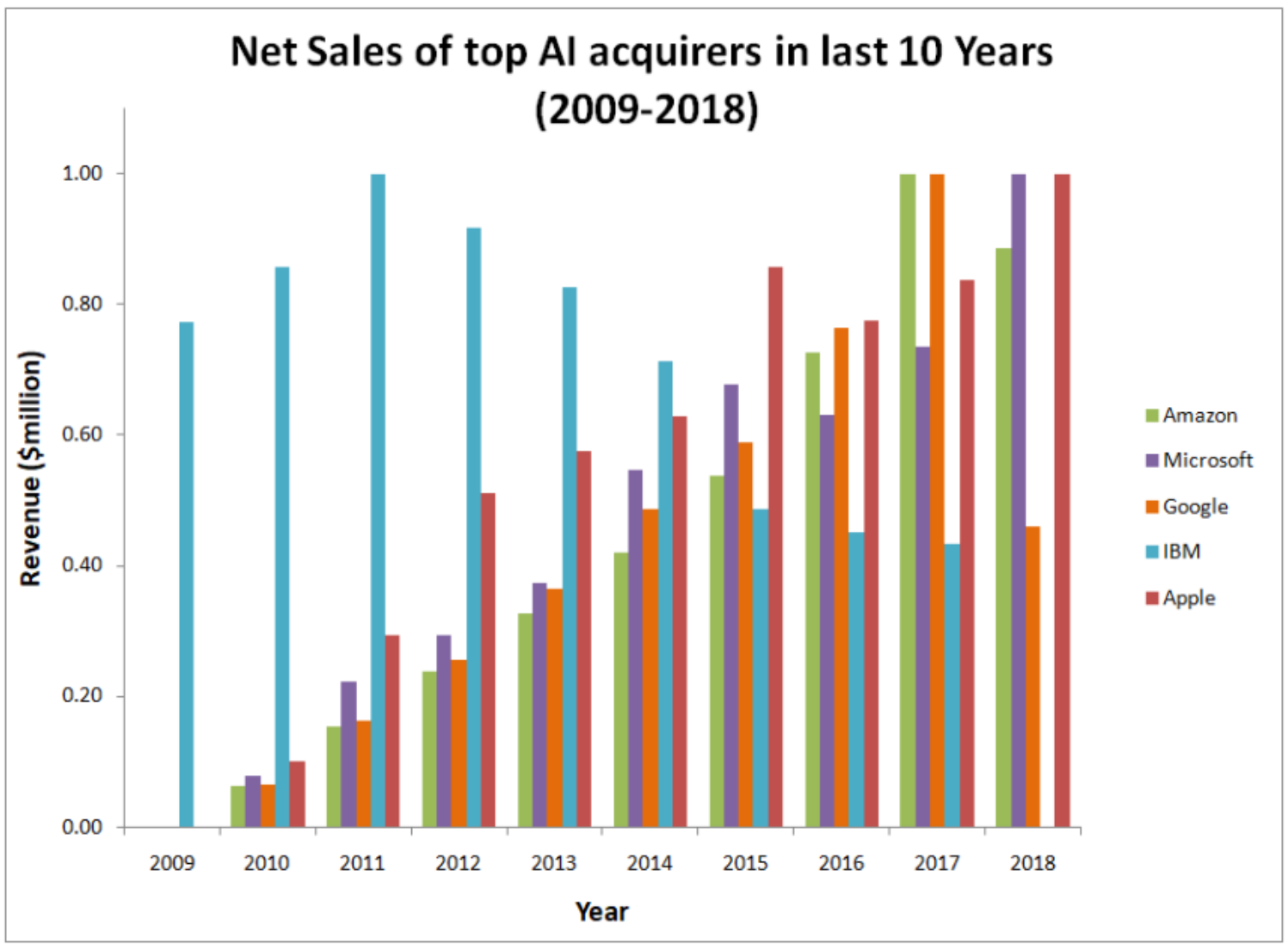

**Fig. 4** Net Sales of top AI acquirers in the last 10 Years (2009-2018)

Since start-ups are considered as innovation and growth drivers of the economy, we believe that their analysis would result in important conclusions, relevant to the detection of the impact of automation and AI on business models. Therefore, we procured the list of preeminent AI start-ups for the year 2017 and 2018 with the help of the CB Insights' Mosaic algorithm[2]. The algorithm identifies the top AI start-ups by evaluating the factors like profile, financing history, tech innovation, patent activity, team strength, investor quality, business model, mosaic score, funding history etc. The two lists were made available by analyzing, 1650+ and 2000+ global start-ups respectively, using the Mosaic algorithm. In the rest of the paper, we will refer to the AI start-ups list for 2017 and 2018 as AI17 and AI18 respectively.

We further analyzed the start-ups listed in AI17 and AI18 to investigate the impact of AI on businesses. The following section lists the investigations and results obtained from the data of AI17 and AI18. First, we identified the different sectors in which these 200 newborn companies have managed to prove themselves as top AI starts-ups. Second, we evaluated the total funding received by these sectors. Third, we determined the geographical distribution of the top AI start-ups; the results guide us towards the exploration of the "Dark side" to AI. A part of these results has been presented at an international conference DIGITS 2018 (Soni 2018a) jointly organized by University of Maryland and Birla Institute of Management and Technology.

[2] Mosaic algorithm: It is a data- driven technology developed by CB Insights to measure the growth of private companies. The algorithm uses machine learning and advanced language processing techniques to understand these companies.

### 5.2.1 Sectors and Industries

In this section, we identified different industries in which AI17 and AI18 develop and/or implement AI technology viz. autonomous vehicles, business intelligence, healthcare, cybersecurity, robotics etc. Fig. 5(a) and (b) depicts the percentage of AI17 and AI18 in different industries, the figure shows all the fields where automation and AI is showing its impact; from home to industry, education to healthcare, the spread of AI is all pervasive. From Fig. 5(a), we can see that in AI17 the majority of the start-ups were concentrated in the industries dealing with core AI, healthcare and business intelligence, while in AI18 (Fig. 5(b)) the focus shifted to cybersecurity, cross-industry and enterprise AI.

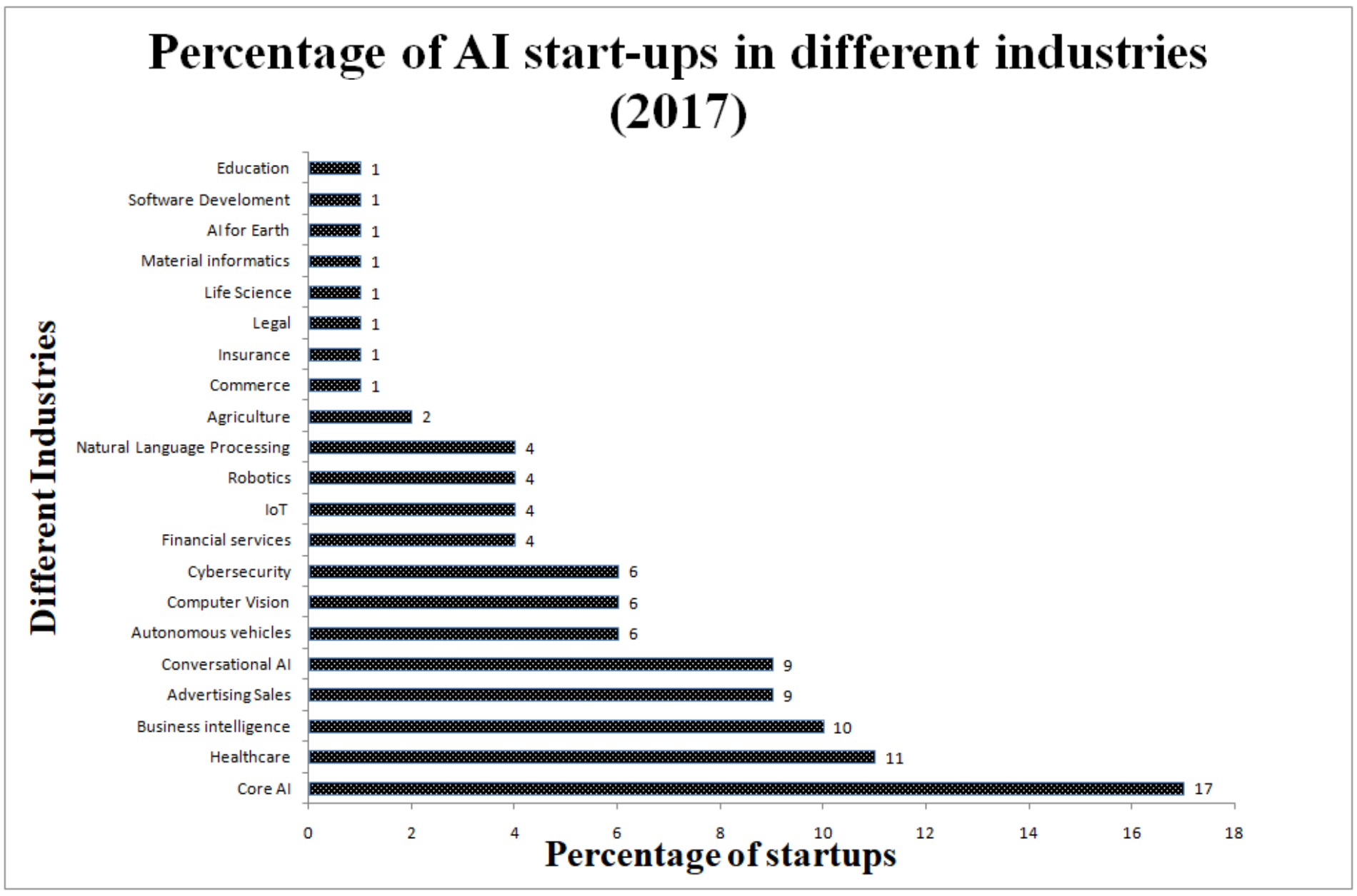

**Fig. 5(a)** Percentage of AI start-ups in different industries (2017)

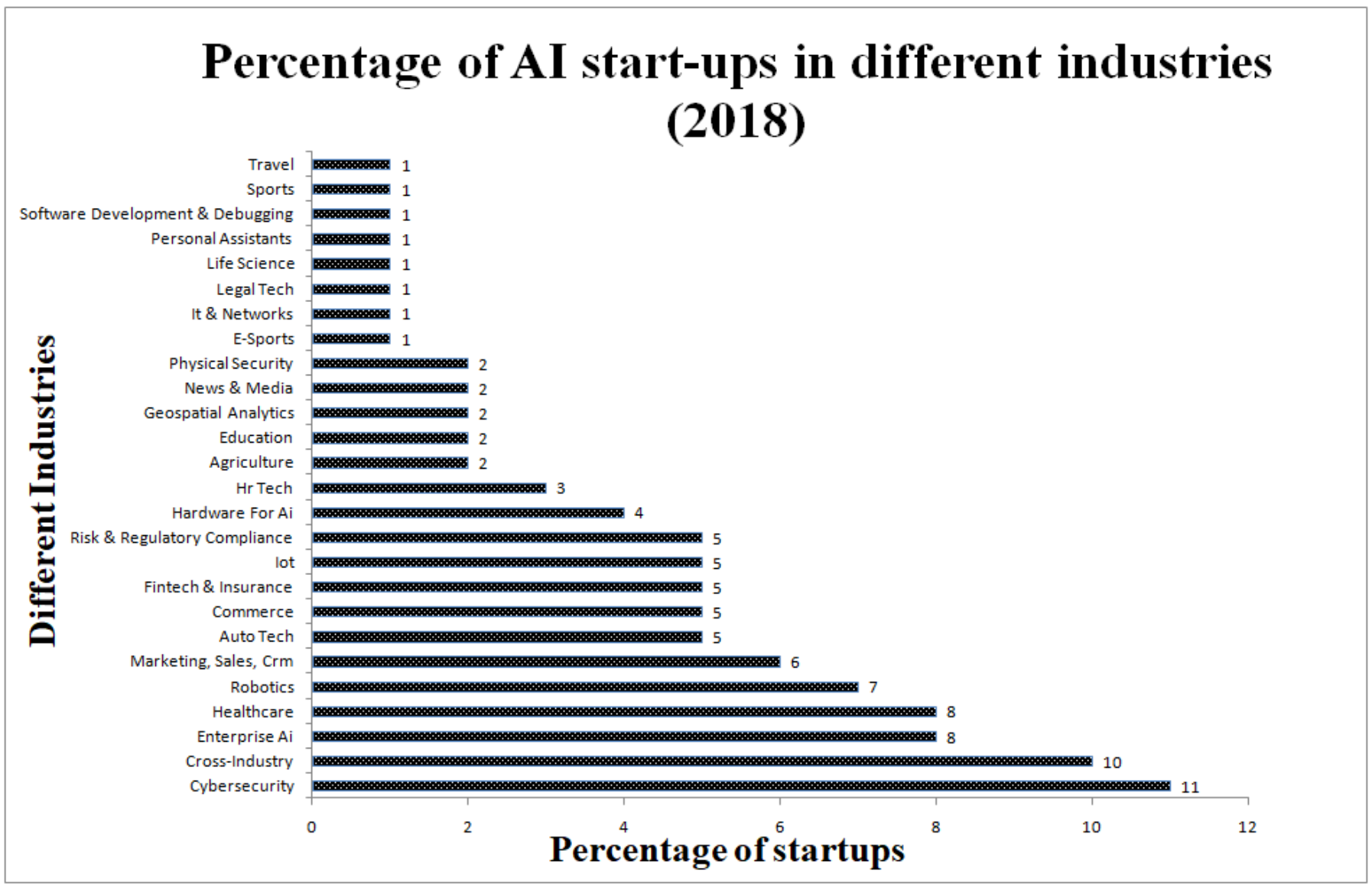

**Fig. 5(b)** Percentage of AI start-ups in different industries (2018)

Fig. 5 summarises a total spread of forty-seven industries (twenty-one for AI17 and twenty-six for AI18) in which top 200 start-ups develop and/or implement AI technology. We identified the eight industries which are common in AI17 and AI18 viz. healthcare, cybersecurity, business intelligence, marketing & sales, autonomous vehicles, financial services, IOT, and robotics. The inclination of start-ups towards these industries indicates that these industries will create more opportunities in near future and provide improved goods and services by enabling the automation of many tasks.

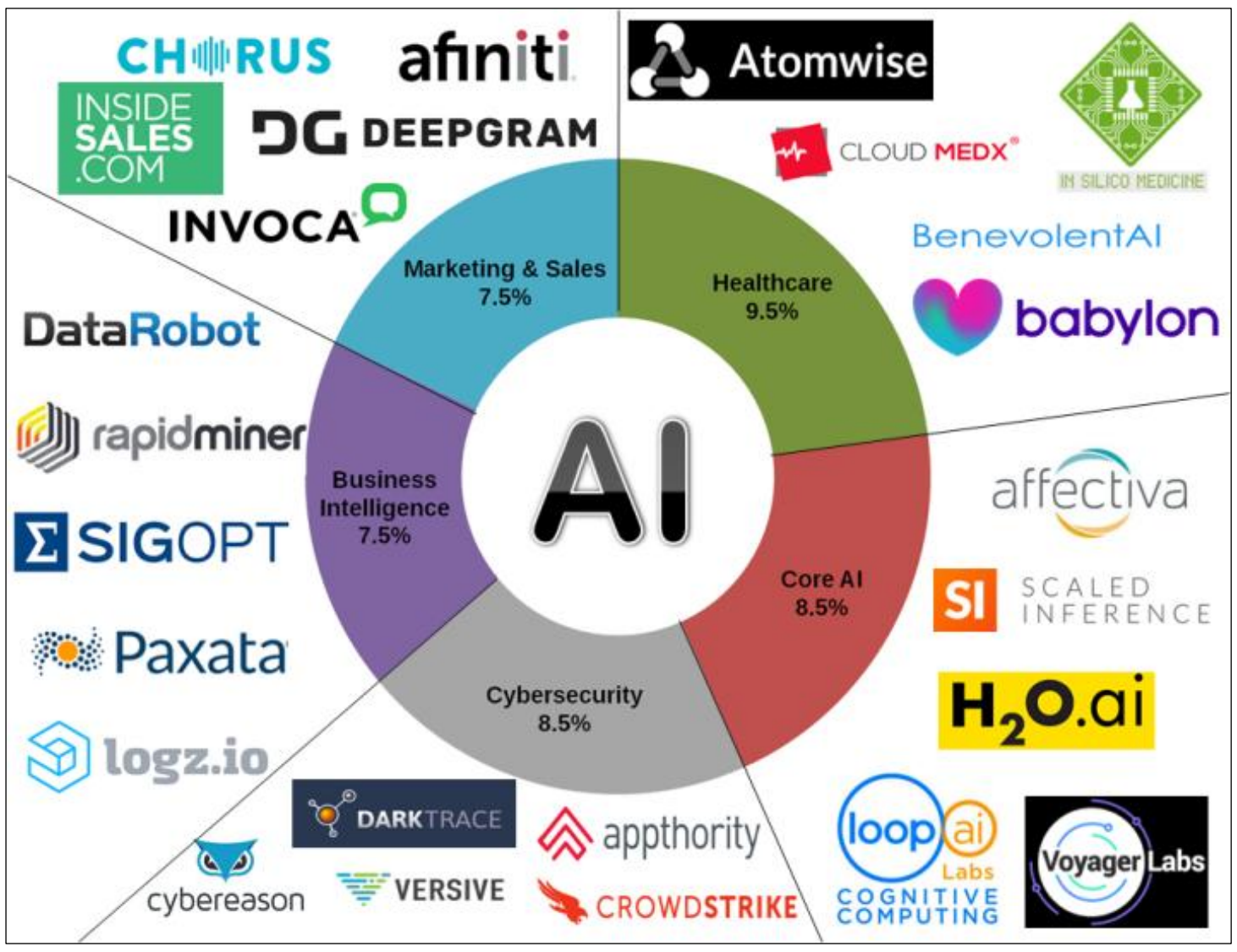

**Fig. 6** Start-ups working in top five industrial sectors (The percentage in the figure depicts the start-ups engaged in the industry out of the total of 200.)

Analyzing the data of AI17 and AI18 industries, we find that the top five industries where AI is maximally used are healthcare, cybersecurity, core AI, business intelligence and marketing & sales (Fig. 6). Some of the start-ups mentioned in the figure have become brand names in their fields in just a few years of their launch. This analysis provides an insight into the upcoming tendency of AI industries.

### 5.2.2 Funding

Next, we investigated the funds received by AI17 and AI18 in the past few years. Fig. 7 depicts the year wise funds raised by the AI17; the total amount received in 2011 was $25.88 million, this increased exponentially by 71.13% in a short span of 6 years resulting in $1866.6 million in 2016. We can also see that there is an almost linear increase in the number of start-ups in the same duration.

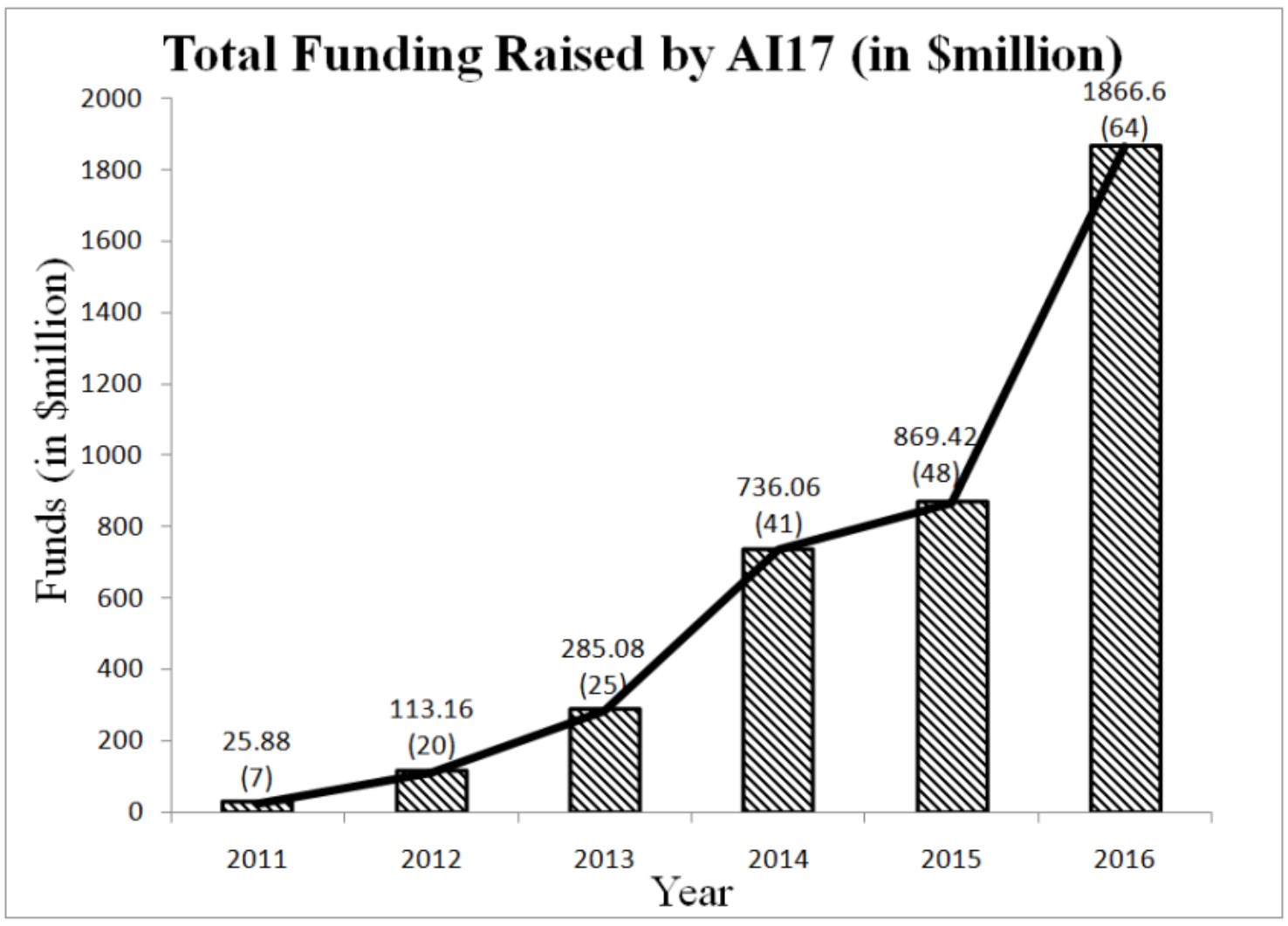

**Fig. 7** Year wise funds (in $million) raised by AI17 (Each bar indicates the total funds received and the number of start-ups received funds in that particular year)

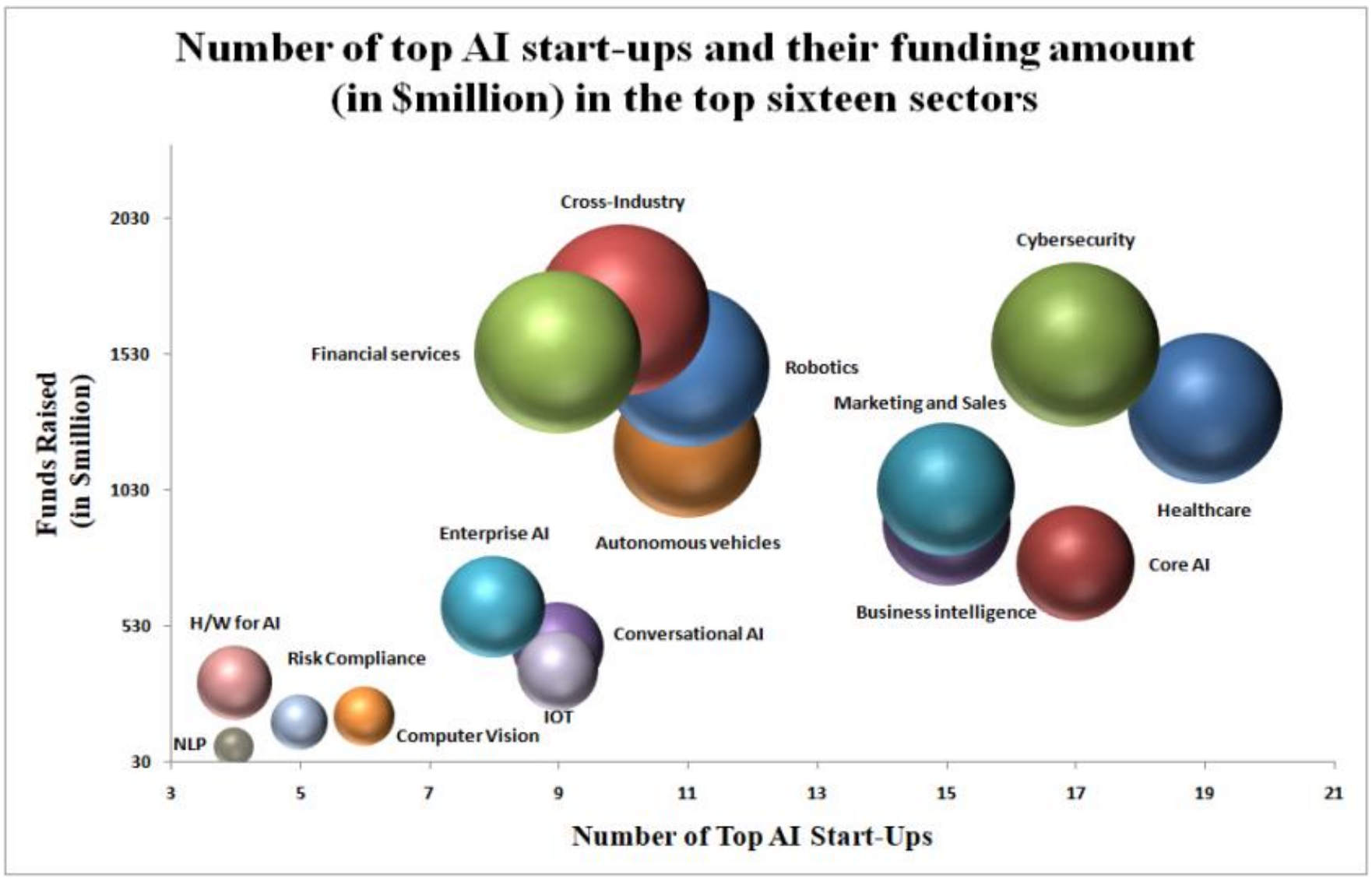

**Fig. 8** Number of top AI start-ups (out of 200) and their funding amount (in $million) in the top sixteen sectors

Additionally, the total funding raised by AI18 is $12.74 billion, this is more than double (2.27 times) the amount raised by AI17. This increment shows the rising interest of investors in AI. Based upon the funding and involvement of the start-up in different industries, we identified top sixteen industries in AI17 and AI18. Fig. 8 illustrates a bubble plot with top sixteen industries wherein the size of the bubble indicates the total funding in millions (USD). The figure clearly shows that cross-industry, cybersecurity, financial services, and robotics are the four industries with maximum funds. These four industries have managed to attract 35% of the total funding in the top 200 start-ups.

### 5.3 The 'Dark side' to AI

In this section, we have investigated important insights from the geographical distribution of AI17 and AI18. It is surprising to know that out of a total of 195 countries in the world, AI17 and AI18 are located only in thirteen countries i.e. top global AI start-ups are located in only 6.6% of the countries on the earth. In these thirteen AI countries, there are only thirty states which are leading this AI revolution. Fig. 9(a) and (b) shows the percentage of AI17 and AI18 in different parts of the world. The U.S. is leading this revolution with the headquarters of approximately three fourth of the total start-ups with the majority in California, Silicon Valley, the heart of AI.

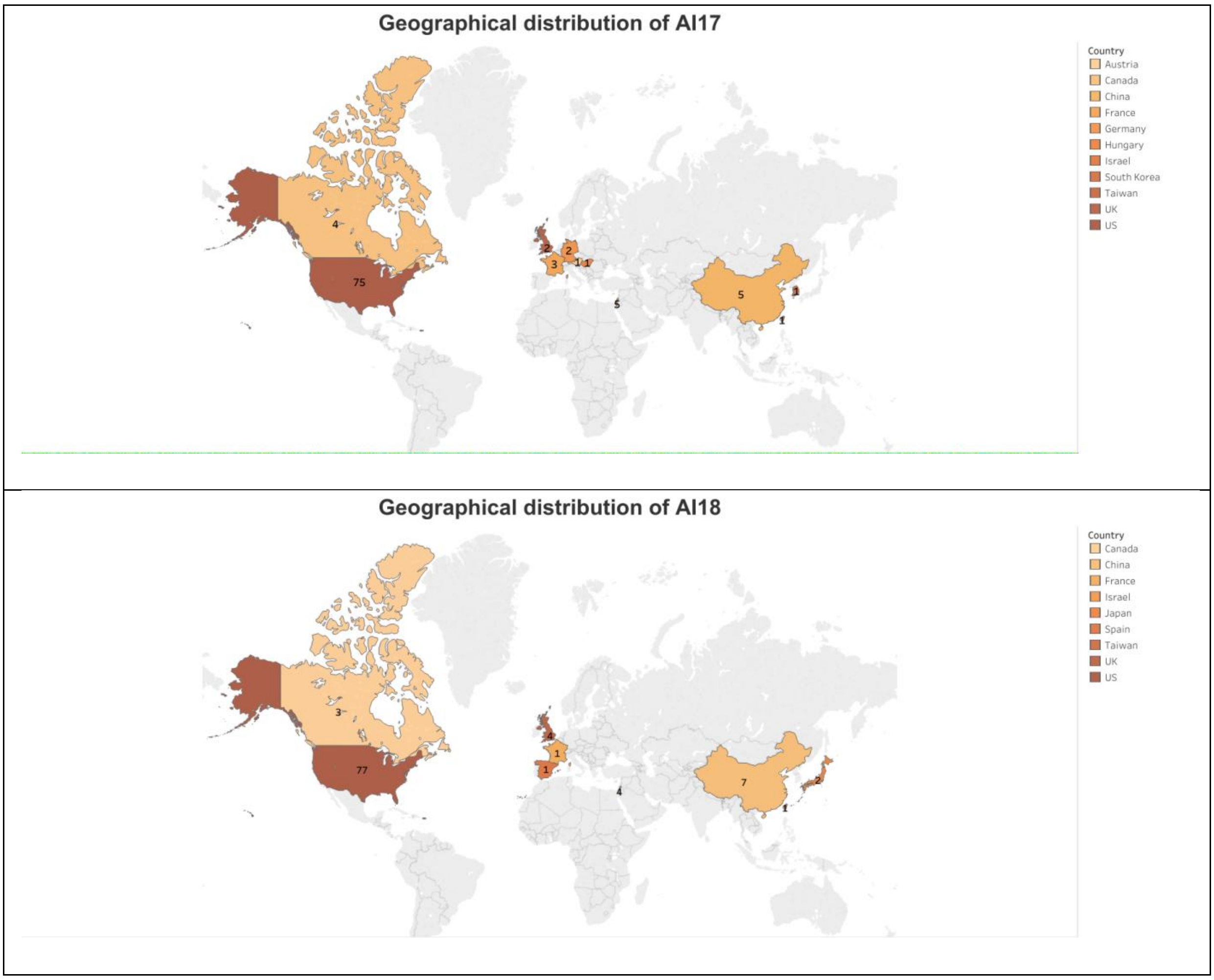

**Fig. 9(a) and (b)** Percentage of AI17 and AI18 in different parts of the world

From the figure above, we can see that the AI talent is confined only to a few regions in the world creating an “AI divide”- the “dark side” to AI. The further expansion of the technology in AI-enabled countries and nonparticipation of other countries will widen this divide. This divide, like the digital divide, would deepen the inequality in economic, cultural and social sectors; would create a chasm. This will have a profound impact on education, income, living standards and businesses. It is imperative that the governments, communities, companies, and individuals should take initiatives to develop their country as an AI country in order to remain technologically independent and stay ahead in the race of AI.

## 6. Shaping of Business Contexts with AI

The adoption of AI technology in the organizations has also led to the shaping of business contexts (factors which influence the performance of the business). We have identified business contexts which have been influenced by AI; we call this the third dimension of our three-dimensional analysis of the overall impact of AI on businesses. For this, we collected data from various sources viz. business press releases, annual reports of the top companies, innovation trend reports and research reports released by the market intelligence firms (Gartner, Forrester, and IDC etc.). We applied the inductive content analysis (i.e. preparation, organization and reporting) (Elo 2008) on the data collected from various sources; the criterion led to the selection of three business contexts where the transformation was considerably visible. In the preparation phase, we captured "themes" (letter, word, sentence or portion of a page) (Elo 2008) from the data; these describe the pervasive impact of AI on the global market (and society) and future predictions. In the organization phase, the captured "themes" were analyzed and organised under different categories, this led us to the identification of three business contexts. After consolidation, we report the three identified categories (business contexts) as Customer Interaction, Sales Platform, and Employee Skill Set.

### 6.1 Customer Interaction

Customer interaction is the most basic form of communication between a company and its customers. Every single interaction is another opportunity for the company to satisfy customers and retain them. In the conventional process, customers used to interact with the employees of the company, holding the position of retail shop executive, salespersons, cashiers, customer relationship managers etc. for all their needs, products and services. The integration of intelligent agents in the companies is transforming the customer interaction from 'human-to-human' to 'human-to-machine'. "Chatbots" and "virtual assistants" are the intelligent conversational agents capable of conducting a human-like conversation with the customers via textual and auditory method respectively. These agents are eliminating delays, human errors and provide almost immediate personalized responses to the customers. A few exemplary companies with the successful implementation of chatbots are:

1. Google duplex: For making real world calls
2. 1-800-Flowers: Order flowers
3. North face: Product selection
4. Spotify: Discover weekly playlist
5. KFC: Facial recognition for order prediction

Currently, the most advanced conversational agents can automate simple, repetitive, low-level tasks and queries. However, developers are on their progressive path to make them capable of performing complex tasks, understand human emotions and thus deliver an efficient and satisfactory/ contented experience to the customers. Several global market intelligence firms have conducted various surveys and performed research activities in order to analyze the present and future of AI-based customer interaction. A few "themes" captured from them have been quoted here-

- Gartner Group, the world's leading research and advisory company, "By 2020, customers will manage 85% of their relationship with the enterprise without interacting with a human." (Marriot 2011)
- Servion, global solution provider, "AI will power 95% of customer interactions by 2025." (Nirale 2018)
- International Data Corporation (IDC), the premier global market intelligence firm, "By 2019, personal digital assistants and bots will execute only 1% of transactions, but they will influence 10% of sales, driving growth among the organizations that have mastered utilizing them." (Fitzgerald 2017)

- Juniper research, digital market research, forecasting & consultancy, “Chatbots will be responsible for cost savings of over $8 billion per annum by 2022, up from $20 million this year.” (Foye 2017)
- Oracle, multinational computer technology corporation, “With regards to chatbots, which are in many ways the most recognizable form of AI, 80% of sales and marketing leaders say they already use these in their CX (customer experience) or plan to do so by 2020.” (Oracle 2016)
- Salesforce, cloud computing company, “By 2021, efficiencies driven by AI in CRM will increase global revenues by $1.1 trillion and ultimately lead to more than 800,000 net-new jobs, surpassing those lost to automation.” (Einstein 2017)

The above “themes” indicate that human-machine interactions will not only provide potential benefits to the customers but are also expected to reduce expenses within the cooperation (Foye 2017). The adoption of AI-based agents will reduce the need for the human force for customer interaction. We agree that this will reduce the number of job opportunities in this area, however, as predicted would also create new job opportunities in new areas (Einstein 2017).

**6.2 The Sales Platform**

The advancement in technology has introduced and encouraged the concept of electronic commerce business models (E-commerce). A number of business organizations have switched from the traditional method to the electronic method to sell their products and services. Automation and AI is strengthening this business model by providing a better buy-sell experience to both buyers and sellers via sales prediction, recommendation engines, warehouse automation, and innovative e-commerce platforms. Amazon, Netflix, Alibaba, and eBay are the brand names for online retailing which have significantly transformed the market through above AI processes. These vendors guide the customers towards the purchase of a product by various means viz. personalized digital advertisements on the web, customized coupons/ offer suggestions, and distinctive emails/ messages. All these tasks make use of past behavior of a customer (as previous purchases, selections or ratings), current search keywords, and many other parameters to recommend different products. Here, we have quoted a few “themes” captured from various market research reports and companies’ website that proves automation and AI is contributing in advancing the e-commerce business platform:

- McKinsey & Company, American worldwide management consulting firm, “35 percent of what consumers purchase on Amazon and 75 percent of what they watch on Netflix come from product recommendations based on algorithms.” (MacKenzie 2013)
- Rakuten, Japan’s largest e-commerce site, “Online shopping businesses usually have hundreds, thousands of FAQs and users need to find the answers from among those. By leveraging AI, we can provide users with the responses they need quickly, 24 hours a day.” (Chatani 2018)
- Forbes, American business magazine, “On Singles' Day (November 11, 2017), Alibaba, the Chinese e-commerce giant, hits $25.4 billion, smashing its own record from last year and cementing it as the world's biggest shopping event. Product recommendation system: Tmall Smart Selection, AI-powered chatbot: Dian Xiaomi and 200 delivery robots are the AI technologies that helped Alibaba and its founder Jack Ma to achieve these records.” (Bernard 2018)

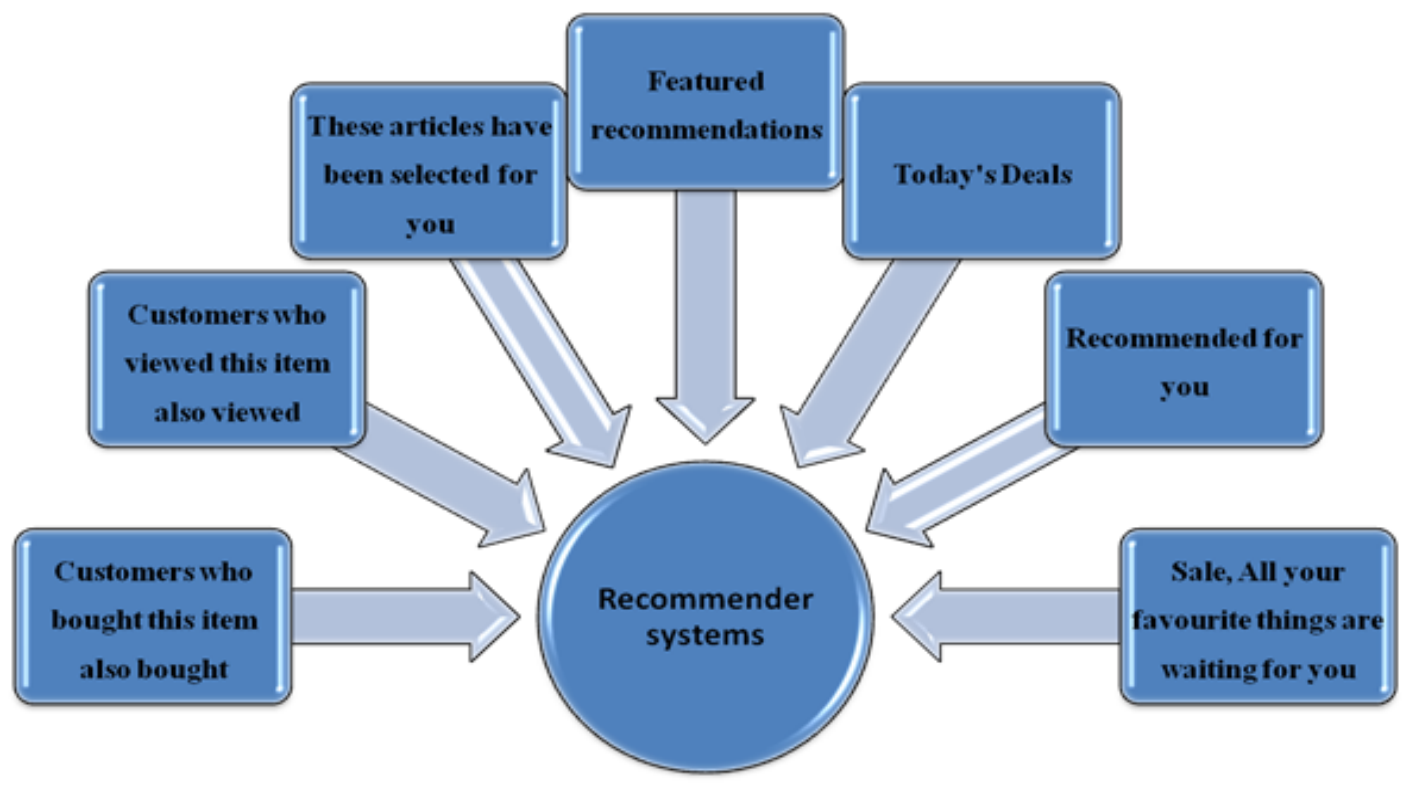

**Fig.10** Phrases generated by different recommender systems

The intelligent empathetic machines personalize the shopping experience of the customer, making him/her comfortable to such an extent where he/she needs not to do much to purchase any product. Some phrases which are generated by different recommender systems, commonly seen on the web or in promotional emails are shown in Fig. 10. The outcomes of these recommender systems may ease the shopping experience of customers. These systems can secure more sales by attracting valuable customers to visit the site more often/ frequently/ again and again. Some of the examples using automation and AI processes in business are product suggestions by retail industry (e.g. Amazon), video suggestion by the entertainment industry (e.g. Netflix and Youtube), songs suggestion (e.g. Spotify and Last.fm), recommendations of grocery products (e.g. Bigbasket), recommendations of books (e.g. Readgeek). In conclusion, the automation and AI systems are transforming the markets from traditional to the digital ones and would continue to revolutionize in near future.

### 6.3 Human Skills

The displacement of jobs has always been a major concern in the process of innovation. Till now, we have seen that every sector wants to grab the phenomenal opportunity to make their sector an AI sector. To fulfill this demand there will be a rise in need of a "new" labour force which will build, develop and productize AI in near future. To gain insights, we made use of the reports released by LinkedIn (a business and employment-oriented service with 500+

million members) in January 2018 and 2019 (Petrone 2018, Petrone 2019, Pattabiraman 2019). In these reports (Petrone 2018, Petrone 2019), the identification of top twenty-five skills was done by analyzing the hiring and recruiting activity on LinkedIn for the previous years. We observed, for the year 2018 out of all the identified top skills, 36% (more than one-third) of the skills are AI-related (Petrone 2018) and similar results have been observed for the year 2019 as shown in Table 4 (Petrone 2019, Pattabiraman 2019). The table below illustrates the details of the top AI-related skills, jobs related to them and approximate salary offered in the US (in USD) (Pattabiraman 2019).

**Table 4** The details of the skills, their related jobs and salary in the US (in USD)

The degree of demand for AI-related skills can be estimated from a recent report released by IBM, Burning Glass Technologies and Business-Higher Education Forum (BHEF) (Markow 2017) and AI Index Steering Committee (Shoham 2018). As per the report (Markow 2017), there will be a 15% increase in the number of job releases by 2020, which makes approximately 2.7 million vacancies in the US alone. According to the report, in the US, some AI-related jobs (data science and analytics jobs) remain open for 5 to 10 extra days after the release of the advertisement. This is in comparison to the average opening days of other job advertisements. The situation is due to the unavailability of the required AI-skilled candidates. This demand and supply problem of data professionals will remain for years as the employers demand at least a master or a Ph.D. degree with an additional 3 years of experience in the related sector (Markow 2017). This is creating a big challenge in recruitment and thus disrupting product development.

Also, according to International Data Corporation (IDC), the premier global market intelligence firm:

- “By 2020, 85% of New Operation-Based Technical Position Hires Will Be Screened for Analytical and AI Skills, Enabling the Development of Data-Centric DX Projects without Hiring New Data-Centric Talent” (Fitzgerald 2017).
- “By 2023, 35% of workers will start working with bots or other forms of AI, requiring company leaders to redesign operational processes, performance metrics, and recruitment strategies.” (Murray 2018)

Thus, AI-related skills are the skills of the future. There is a need for the identification and implementation of strategies for the preparation of upcoming future. We recommend a few initiatives that can be taken to create a large pool of skilled AI-professionals. At fundamental level, students can be exposed to mandatory data-focused programs: machine learning and AI courses. Data labs can be established at educational institutes. Countries like China have already started fostering AI education at their schools and universities (Xinhua 2018). To create a strong data environment and leverage AI most efficiently, country leaders, department and technology heads should be given AI primers; this will make them aware of the capabilities and potential benefits of AI-based machines and services. Attention is needed in closing the gap between the higher education and the industrial requirements. To address this issue, countries should encourage the investigations which identify the shifts in market demand (Sousa 2019) and equip employees with relevant professional competencies such as training courses in areas viz. data mining, machine learning, business intelligence, basic statistics and deep learning, analytical reasoning, machine handling.

## 7. Conclusion remarks

We can see that AI is not hype but has the capability of transforming the global economy via technological innovations, scientific knowledge and entrepreneurial activities. The progressive growth of automation and AI in the last decade is attributed to two major factors: the increasing availability of big data and hardware accelerators (GPUs and TPUs). These factors are making AI the core technology responsible for extreme automation and connectivity and thus, taking the world towards the dawn of the fourth industrial revolution. This will have profound impacts on governments, communities, companies, and individuals.

The extreme high capability of intelligent agents (IA) in various games, recognition and classification tasks offer opportunities for process innovation as well as product innovation. This is leading to the development of assistive technologies and products for the disable and elderly people. This is also advancing the toy and gaming industry which will enhance the entertaining experience and develop cognitive & emotional intelligence of children. Conclusively, the involvement of the autonomous technology in almost every sector and launch of a large number of AI-based machines and services would improve health, educational opportunities, security, transportation, safety, trade and every other aspect of living. However, there are some security, privacy and ethical concerns associated with the use of AI technology which needs a lot of attention.

The innovation process and global competitiveness is strengthening as an outcome of the adoption of various strategies by the corporate firms (companies and start-ups) to become AI-firms. The actual intention is to grow with the most advanced technology of AI and win the technological race. Our study uncovers the top automation and AI industries that will create more opportunities in near future viz. healthcare, cybersecurity, core AI, business intelligence and marketing & sales. The upward trajectory in the investment in automation and AI in the last 8 years clearly implies that it has the potential to change the economy. However, we can see that the AI is confined only to a few regions in the world creating an “AI divide”. The further expansion of the technology in AI-enabled countries and nonparticipation of other countries will widen this divide. This divide, like the digital divide, would deepen the inequality in economic, cultural and social sectors; would create a chasm. We see this divide as the “dark side” to AI.

Our three-dimensional analysis has provided the state-of-the-art of research, innovation and market deployment of automation and AI. It has also provided a better understanding of how AI can reshape the markets, transform the innovation processes, the organization of research & development, business processes and the global economy. The knowledge about all these domains will make the actors aware about the development and adoption of AI in near future. This will also aid the policymakers in identifying challenges and exploring the legal and ethical corners in the field of AI. Conclusively, AI will have a profound positive impact on education, living standards, health and businesses but, initiatives must be fostered towards the exploration of ethical issues and reducing the AI divide among countries.

**Table 1** Commercially available intelligent machine and services, the intelligent software behind them, their developers and their user base.

| Task | Intelligent Software/ Device | Developer | Compatible Products/ Capabilities | Number of Users |
|---|---|---|---|---|
| Voice-based virtual assistant | Alexa | Amazon | Speaker, light bulbs, switches, lamps, and other home automation products | 100 Million (January 2019) |
| | Siri | Apple | Watch, computer, mobile, speaker, television, lights, fans, switches, locks, cameras, windows, and other home automation products | 500 Million (January 2018) |
| | Google Assistant | Google | Speaker, refrigerator, oven, car, washer, and dryer | 500 Million (May 2018) |
| | Cortana | Microsoft | Windows 8.1 and 10 computers | 700 Million |
| | | | and mobiles, watch, video game, glasses, headsets, speaker | (June 2018) |
| Web Mapping | Google Maps | Google | Android and iOS-based devices | 1 Billion (May 2018) |
| Ridesharing Apps | Uber | Uber.com | Android and iOS-based devices | 75 Million (December 2018) |
| | Lyft | Lyft.com | Android and iOS-based devices | 23 Million (January 2018) |
| Filter Spam | DeepText | Facebook | Can provide simple suggestions like offering Uber or a Lyft, selling or purchasing options etc by understanding text on messenger. | 2.23 Billion (June 2018) (Currently, processing 10000 posts/ second in 20 different languages) |
| Humanoid Robots | Pepper | SoftBank Robotics | • Optimized for human interaction<br>• Can engage with people through conversation and touchscreen. | 2000 companies (October 2018) |
| Healthcare | Virtual Medical | Sense.ly | Blood pressure machine, glucose meter, weight machine | Not disclosed |
| | Assistant (Molly) | | | |
| | Disease Diagnosis- IBM Watson | IBM | Identify key information in a patient's medical record and explore treatment options. | Adopted by various hospitals and cancer centres worldwide. |
| Collaborative robot (Cobot) | Robot CR | FANUC | Can recognize and lift heavy objects | Not disclosed |
| | Kiva Robot | Amazon Robotics | Capable of automating the picking and packing process in industries. | 15000 (at Amazon warehouses in the U.S.) |
| Self-Driving Cars | Nissan NV200 | Drive.ai | Can navigate without human intervention | In the initial phase of testing on real roads |
| Assistive Device for Blind | Horus | Eyra | It consists of a headband and a stereo camera that can recognize text, faces, and objects. | In the initial phase of testing |

**Table 2** The details of the acquisitions and NIPS sponsorship level of the top five AI companies.

| S.No. | Top Acquirers | Sponsorship | Acquisitions | Number of acquisitions |
|---|---|---|---|---|
| 1. | Google | Platinum | Terraform Labs Incorporated, Superpod, Inc., Bitium, Inc., AIMatter, Halli Labs, Api.ai, Moodstocks, Timeful, Granata Decision Systems, Jetpac, Emu Messenger, DeepMind Technologies, DNNresearch, CleverSense, Vision Factory, Dark Blue Labs, Nest Labs, Waze, PittPatt, Numovis, Wavii, Kaggle, Kifi, Tenor Inc, Fabrics, Eyefluence, Urban Engines, Apigee, Revolv, Bot & Dolly, Holomni, Meka Robotics, Redwood Robotics, Industrial Perception Inc., Behavio, PostRank, Phonetic Arts. | 37 |
| 2. | Apple | Platinum | Silk Labs, Inc., Asaii, Siri, SensoMotoric, Regaind, Init.ai, Pop Up Archive, Lattice, RealFace, tuplejump, Turi, Emotient, Perceptio, Vocal IQ, Novauris Technolgies, PolarRose, Cue, Concept.io, Flyby Media, Shazam Entertainment Limited, Gliimpse Inc., faceshift AG, Semetric Limited, Acunu Ltd., PrimeSense Ltd. | 25 |
| 3. | IBM | Diamond | Armanta Corporation, Cloudigo, Agile, EZ Source, Blue Wolf Group LLC, Truven Health Analytics, Iris Analytics, | 21 |
| | | | The Weather Company, Merge Healthcare Inc., Explorys, AlchemyAPI, Cognea, Silverpop Systems Inc., Star Analytics, StoredIQ, Butterfly Software Ltd., Tealeaf, i2 Limited, Netezza, Coremetrics, Expert Personal Shopper (XPS) | |
| 4. | Amazon | Platinum | Harvest.ai, Angel.ai, Orbeus, Evi Technologies, Sqrrl, Ivona, Kiva systems, Blink, Body labs, Graphiq, Yap, Ring, Safaba translation solutions, 2lemetry, TenMarks Education Inc., Goodreads, Amiato, CloudEndure Ltd, TSO Logic | 19 |
| 5. | Microsoft | Diamond | Bonsai, GitHub Inc., Semantic Machines, AltspaceVR, Maluuba, Genee, Solair, Metanautix, VoloMetrix Inc., Equivio, Aorato, Netbreeze, id8 Group R2 Studios, MarketingPilot, TouchType Ltd., Lobe | 16 |

**Table 3** The details of the skills, their related jobs and salary in the US (in USD)

| S. No. | Skills | Related Jobs | Approximate US Salary (in USD) |
|---|---|---|---|
| 1 | Machine Learning, Python, Data Mining, Artificial Intelligence, Data Science | Machine Learning Engineer | 182000 |
| 2 | Data Science, Data Mining, Data | Data Scientist | 130000 |
| | Analysis, Python, Machine Learning | | |
| 3 | Analytics, Software Development, Business Analysis, | Engagement Manager, Scrum Master, Information Technology Lead, Product Owner | 113750 |
| 4 | Cloud Computing, Enterprise Software, Software-as-a-Service, Business Analysis | Solutions Consultant | 110000 |
| 5 | Cloud Computing, Software Development, Amazon Web Services, Solution Architecture, Linux, Python | Cloud Architect, Solutions Architect, Site Reliability Engineer | 107750 |

**Acknowledgment.**

This work is sponsored by Department of Science & Technology, Ministry of Science & Technology, New Delhi, India.